\documentstyle[preprint,tighten,floats,prd,aps]{revtex}
\def \DESepsf(#1 width #2){\bf #1  here: just uncomment the macro.}
\begin{document}
\preprint{OITS 545}
\draft

\title { Top quark contribution to hadronic decays\\
 of the $Z$-boson at $\alpha_{s}^{2}$ in QCD}
\author{Davison E.\ Soper and Levan R.\ Surguladze}
\address{Institute of Theoretical Science,
University of Oregon, Eugene, OR  97403}
\date{27 June 1994}
\maketitle
\begin{abstract}
We evaluate the effect of a virtual top quark on the coefficient of
$\alpha_{s}^{2}$ in   the decay rates $\Gamma(Z \rightarrow {\rm
hadrons})$ and $\Gamma(Z \rightarrow b\overline{b} )$. We treat the
dependence on the top quark mass exactly instead of using a large mass
expansion.  The present work completes the evaluation of the
$\alpha_{s}^{2}$ contributions to these quantities. The calculation
uses both the $\overline{\rm MS}$ and Collins-Wilczek-Zee
renormalization prescriptions. The results can be applied to the
hadronic decays of the $\tau$-lepton.
\end{abstract}

%****************************************************

One crucial test of quantum chromodynamics (QCD) is based on the
total cross section for $e^{+}e^{-} \rightarrow {\rm hadrons}$ at
LEP \cite{rev0}. The theoretical expression for
\begin{equation}
R_Z = \Gamma(Z^0 \to {\rm hadrons})/
\Gamma(Z^0 \to e^+e^-)
\end{equation}
is compared to the measured quantity to extract a value for
$\alpha_s(M_Z)$.  If QCD is correct, this value of $\alpha_s(M_Z)$
should be the same as the value obtained in other experiments, such as
deeply inelastic scattering and three jet production in $e^{+}e^{-}$
annihilation. A high precision experiment is required in order to
obtain good precision on $\alpha_s(M_Z)$ because $\alpha_s(M_Z)$ is
small and $R_Z$ has the form
\begin{equation}
R_Z = R_0\left\{1 + (\alpha_s/\pi)+ \cdots \right\}.
\label{RZ}
\end{equation}
Despite this experimental difficulty, the use of this measurement as a
QCD test is regarded as ``gold plated'' because of the theoretical
simplicity of $R_Z$. Simple analyticity arguments relate $R_Z$ to the
behavior of the Z-boson propagator at short distances, without the need
for complicated arguments about factorization of long-distance effects,
or about the infrared safety of jet definitions.  To be sure, there are
``hadronization effects,'' but at $M_Z \approx 90\ {\rm GeV}$, these are
easily seen to be negligible.

The test may also be regarded as ``gold plated'' because calculations
for $R_Z$ are quite complete. The leading contributions in the
approximation that the $u$, $d$, $s$, $c$, and $b$ quark masses vanish
while the $t$ quark mass is large compared to $M_Z$ are known up to
order $\alpha_s^3$ \cite{4loopV,4loopA,3loopV,3loopA}. Corrections
to this limit, including the complete dependence on $m_t$, are known
for most graphs up to order $\alpha_s^2$ \cite{3loopA,newchet}. The
one exception concerns the $\alpha_s^2$ graphs shown in
Fig.~\ref{ourgraphs}, in which there is a top quark loop. Here, the
result is known only in an expansion about $m_t/ M_Z =\infty$\
\cite{Chet}. In the real world, it appears that $m_t$ is not
much greater than $M_Z$ \cite{CDF}, so that it is not evident how good
this approximation is. The purpose of this paper is to fill in this
gap by calculating the graphs of Fig.~\ref{ourgraphs} for arbitrary
$m_t/M_Z$.

\begin{figure}[htb]
\centerline{ \DESepsf(ourgraphs.epsf width 10 cm) }
\caption{Contributions to the Z-boson propagator with
a heavy quark loop (represented by the thick line).
The weak current operators are represented by crosses.
The discontinuity of these diagrams contributes to $R_Z$.}
\label{ourgraphs}
\end{figure}

Our motivation for undertaking a calculation of what is, in fact, a
small correction to $R_Z$ was twofold. First, we wanted to know for
sure that the correction is small. Second, we wanted to develop
methods for calculating multiloop Feynman graphs with non-zero
masses. There are powerful techniques that are used for multiloop
massless graphs, but these techniques cannot be applied when masses
are present. One can perform expansions in powers of $m$ or $1/m$,
as noted above, but when $m$ is actually neither very large nor
very small compared to the momentum scale of the problem, it would
be better to be able to include $m$ in the calculation without
approximation. We anticipate applications of the techniques and
results used in this paper to other problems, especially the
calculation of the hadronic decay width of the $\tau$-lepton. For the
case of $Z$-boson decay, we have in mind that the heavy quark in
Fig.~\ref{ourgraphs} is the top quark, but, since the mass is
arbitrary, the results may also be applied to bottom or charm quark
loops.

In order to state clearly what contribution to $R_Z$ we are
calculating, we begin the exposition by outlining the theoretical
structure of $R_Z$ that is relevant for phenomenology. We define the
perturbative expansion coefficients of $R_Z$ by
\begin{equation}
R_Z = \sum_{n=0}^\infty
\left({ \alpha_s(N_L,\mu) \over \pi}\right)^{\!n}
R_n\,.
\end{equation}
Here $\alpha_s$ is evaluated at scale $\mu$, as defined in the
$\overline{\rm MS}$ renormalization scheme with $N_L$ light quark
flavors. Here we may take a ``light'' flavor $f$ to be one for which
$m_f <\sqrt s /2$. For $\sqrt s \approx 90 {\rm \ GeV}$, the number of
light flavors is $N_L = 5$. At order $\alpha_s^0$, the structure of $R$
is simple:
\begin{equation}
R_0 = \sum_{f\le N_L}\ \sum_{i = V,A}
{\cal R}_f^i\!\left(m_f^2(\mu)/s\right)\,.
\label{R0}
\end{equation}
Here there is a sum over the light flavors $f$. There is also a sum
over an index $i = (V,A)$ that labels the two contributions, vector,
$\langle J_V^\mu J_V^\nu\rangle$, and axial vector, $\langle J_A^\mu
J_A^\nu\rangle$. The corresponding contribution to $R_Z$ at lowest
order in the electroweak interactions and zeroth order in the strong
interactions is denoted ${\cal R}_f^i$. These contributions are
functions of $m_f^2/s$. For the sake of definiteness, we choose the
$\overline{\rm MS}$ definition of the quark mass $m_f(\mu)$.  Then the
mass depends on the renormalization scale $\mu$.  The values of the
${\cal R}_f^i$ may be found, for instance, in Ref.~\cite{Born}.

In the Born term (\ref{R0}) it is convenient to keep the exact
dependence on the quark masses. For the QCD corrections, an
expansion about $m_f^2/s = 0$ suffices since, in fact,
$m_f^2/s \ll 1$ for the heaviest ``light'' quark, the $b$. Thus
we write the first order corrections as
\begin{equation}
R_1=
\sum_{f\le N_L}\ \sum_{i = V,A}
{\cal R}_f^i\!\left(0\right)
\Biggl[\ 1
+ \left(m_f^2\over s \right)
\biggl( A_{1,0}^i + A_{1,1}^i \log\!\left(\mu^2/ s\right)\biggr)
\Biggr]
+{\cal O}(m_L^4/s^2).
\label{R1}
\end{equation}
Here $A_{1,0}^V = 12$, $A_{1,1}^V = 0$, $A_{1,0}^A = -22$, and
$A_{1,1}^A = -12$. The value of $A_{1,1}^i$ follows simply
from the renormalization group applied to the $\mu$ dependence of
$m(\mu)$ in $R_0$; the $A_{1,0}^i$ are easily calculated and may
be found in Ref.~\cite{newchet}. The notation $+{\cal O}(m_L^4/s^2)$ is
intended to indicate that we have neglected terms that are no larger
than $m_L^4/s^2$, where $m_L$ denotes the heaviest of the light quark
masses, times constants and logarithms of $s$, $\mu$, and the quark
masses.

At order $\alpha_s^2$ we have
\begin{eqnarray}
R_2&=&
\sum_{f\le N_L}\ \sum_{i = V,A}
{\cal R}_f^i\!\left(0\right)
\Biggl[
1.985707 - 0.115295 N_L + \beta_0(N_L) \log\!\left(\mu^2 / s\right)
\nonumber\\ &&\hskip 2 cm
+ \left(m_f^2\over s \right)
\biggl( A_{2,0}^i + A_{2,1}^i \log\!\left(\mu^2/ s\right)
+ A_{2,2}^i \log^2\!\left(\mu^2/ s\right)\biggr)
\nonumber\\ &&\hskip 2 cm
+\sum_{f' \le N_L} F(m_{f'}^2/s)
+\sum_{f' > N_L} G(m_{f'}^2/s)
\Biggr]
\nonumber\\ &&\quad
+ {\cal O}({m_L^4/ s^2})
+ {\cal O}({m_L^2/ s})
\times {\cal O}({s/m_H^2})
\nonumber\\ &&\quad
+{\cal L}
\,.
\label{R2}
\end{eqnarray}
Here the first line represents the result with $N_L$ flavors of
massless quarks, including all order $\alpha_s^2$ graphs except that
shown in  Fig.~\ref{triangle}. The presence of the term $\beta_0
\log(\mu^2/s)$, where $\beta_0(N_L) = (33 - 2 N_L)/12$, follows from
the application of the renormalization group to the $\mu$ dependence of
$\alpha_s(\mu)$ in $R_1$. The second line gives the first
corrections due to the non-zero mass of the quark $f$ to which the
Z-boson couples. The values of $A^i_{2,1}$ and $A^i_{2,2}$ follow by
direct calculation or by applying the renormalization group, while the
$A^i_{2,0}$ follow from direct calculation and can also be found in
Ref.~\cite{newchet}:
\begin{equation}
\begin{array}{ll}
A_{2,0}^V = 253/2 - 13 N_L/3 \ \ \ &
A_{2,0}^A = -108.140 + 4.4852 N_L
\\
A_{2,1}^V = 57-2 N_L &
A_{2,1}^A = -155 + 16 N_L/3
\\
A_{2,2}^V = 0 &
A_{2,2}^A = -57/2 + N_L\,.
\end{array}
\end{equation}

In the third line, we encounter the functions $F$ and $G$ that are the
subject of this paper. Their presence is due to the graphs in
Fig.~\ref{ourgraphs}.  In these graphs, a quark of flavor $f'$
appears in an internal loop. Each {\it light} quark
$f'$ ({\it i.e.}\ $u,d,s,c,b$ for $\sqrt s \approx 90\ {\rm GeV}$)
makes a contribution $-0.115295 + F(m_{f'}^2/s)$. Here, by definition,
$F(0)=0$. Each {\it heavy} quark ({\it i.e.}\ top) makes a contribution
$G(m_{f'}^2/s)$. The decoupling theorem \cite{DecTheor} guarantees
that $G$ vanishes as $m_{f'}^2/s \to \infty$. The first term in an
expansion of $G$ about $m_{f'}^2/s =\infty$ was given in
Ref.~\cite{Chet}. Our object is to calculate $G$ for $m_{f'}$ not
much bigger that $\sqrt{s}$. The calculation of $G$ yields a
calculation of $F$ as a corollary. In calculating $F$ and $G$, we
take the mass $m_f$ of the quark in the outside loop to be zero. Thus
in Eq.~(\ref{R2}) we neglect terms from graphs with light quarks only
that are of order $m_L^4/ s^2$ times logarithms.  In graphs like
those of in Fig.~\ref{ourgraphs} with a light quark on the outside
loop and a heavy quark on the inside loop, we neglect terms of order
$m_L^2/ s$ times $s/m_H^2$ times logarithms, where $m_H$ is the
lightest heavy quark mass ({\it i.e.}\ $m_t$).

We have yet to mention the term $\cal L$ in Eq.~(\ref{R2}), which
corresponds to the graph shown in Fig.~\ref{triangle}. In the other
graphs contributing to Eq.~(\ref{R2}), top quark contributions
vanish if one takes the limit $m_t \to \infty$. This does not happen
here because the top quark is required in order to
cancel the contribution of the bottom quark to the triangle anomaly.
These graphs are calculated in Ref.~\cite{3loopA} for arbitrary
$m_t$ with $m_u = m_d = m_s = m_c = m_b = 0$. For $m_t \gg
\sqrt s$ the result has a simple expansion \cite{3loopA},
\begin{equation}
{\cal L}= {\cal R}_b^i\!\left(0\right)
\left\{
\log\frac{m_t^2}{s}
- 3.083
+ 0.0865\,\frac{s}{m_t^2}
+ 0.0132\biggl(\frac{s}{m_t^2}\biggr)^2
+ \cdots
\right\}.
\label{eq:L}
\end{equation}
Corrections to Eq.~(\ref{eq:L}) proportional to $m_b^2/s$ are given in
Ref.~\cite{newchet}.

\begin{figure}[htb]
\centerline{ \DESepsf(triangle.epsf width 10 cm) }
\caption{Graph for $\cal L$, Eq.~(\protect\ref{R2}).}
\label{triangle}
\end{figure}

In the order $\alpha_s^3$ term, $R_3$, it suffices for practical
purposes to simply set $m_L = 0$, and keep only terms that do not
vanish as $m_H \to \infty$. The large $m_H$ form of those graphs
that are order $\alpha_s^3$ corrections to ${\cal L}$ has
been calculated in \cite{4loopA}, and the rest of $R_3$ has been
calculated in the $m_L = 0$, $m_H = \infty$ limit in \cite{4loopV}.
These terms amount to about a 1\% correction in the value of
$\alpha_s$ extracted from
$R_Z$.

Our aim is to calculate the functions $F(m^2/s)$ and $G(m^2/s)$.
To begin, we notice that these functions are related. To see this,
consider $R_Z$ at orders $\alpha_s^0$, $\alpha_s^1$, and
$\alpha_s^2$, ignoring the contribution $\cal L$. Focus
attention on the contributions from a particular choice $i
= V\ {\rm or}\ A$ and from a particular light flavor $f$ in the
outside loop. For our present purposes, it suffices to take $m_f =
0$ and $\mu = \sqrt s$. Call the corresponding function $\delta
R_f^i$:
\begin{eqnarray}
\delta R_f^i&=&{\cal R}_f^i
(0)\Biggl\{ 1 + { \alpha_s(N,\sqrt s) \over \pi}
\nonumber\\
&&+ \left({ \alpha_s(N,\sqrt s) \over \pi}\right)^2
\biggl[
1.985707 - 0.115295\, N +
\nonumber\\
&& \hskip 2cm +
\sum_{f' \le N} F(m_{f'}^2/s)
+\sum_{f' > N} G(m_{f'}^2/s)
\biggr]\Biggr\}.
\label{NLchange}
\end{eqnarray}
We have made explicit the fact that the definition of the strong
coupling $\alpha_s$ depends on the number of flavors, $N$, that are
fully included in loops using the $\overline{\rm MS}$ prescription.
This number $N$ is normally taken to be the number of light flavors
$N_L$, as defined by the condition $m_{f'} < \sqrt s /2$. However,
this is a convention that could be varied. What happens if we change
conventions and consider the heaviest of the light quarks play the
role of a heavy quark in Eq.~(\ref{NLchange})? Let the flavor index
of this quark be $f' = q$. Then the explicit factor of $N$ decreases
by 1 and the value of the strong coupling changes,
\begin{equation}
{\alpha_{s}(N,\sqrt s) \over \pi} =
{\alpha_{s}(N-1,\sqrt s) \over \pi}
 -{1 \over 6}\biggl({\alpha_{s}(N-1,\mu) \over \pi}\biggr)^2
         \log\!\left({m_q^2 \over s}\right)
+O(\alpha_{s}^3) .
\label{eq:alphastransform}
\end{equation}
Furthermore, we replace $F(m_q^2/s)$ by $G(m_q^2/s)$. However the
contribution $\delta R_f^i$ to the physical $Z$-width must remain the
same up to corrections of order $\alpha_s^3$. This requires that
\begin{equation}
G\biggl({m_q^2 \over s}\biggr) =
F\biggl({m_q^2 \over s}\biggr)
 -{1 \over 6}\log\left({m_q^2 \over s}\right)
-0.115295.
\label{FtoG}
\end{equation}

We will utilize this relation by calculating both $F$ and $G$
independently and verifying that Eq.~(\ref{FtoG}) is satisfied. Notice
that, from its definition, $F(m^2/s) \to 0$ when $m \to 0$, while the
decoupling theorem guarantees that $G(m^2/s) \to 0$ when $m \to
\infty$. For small $m$, it is $F$ that appears in Eq.~(\ref{R2}),
while for large $m$ it is $G$.

We now describe the calculation of $F(m^2/s)$. It suffices to
consider the contribution from vector currents $j^{\mu} =
\bar{q}_{f}\gamma^{\mu}q_{f}$, where $f$ denotes one of the light
flavors, the mass of which we take to be zero. The derivation for
axial vector currents is similar, and the results for $F$ and
$G$ are the same. We define the function $\Pi(q^2)$ in terms of the
two-point correlation function of the two vector currents:
\begin{equation}
 i\int e^{iqx}<0|Tj^{\mu}(x)j^{\nu}(0)|0>d^4x =
(q^{\mu}q^{\nu} - g^{\mu\nu}q^2) \Pi(q^2)\,.
\label{pifunctionv}
\end{equation}
On the right hand side of Eq.~(\ref{pifunctionv}), one can separate
the transverse tensor factor because of current conservation. The
corresponding contribution to the hadronic decay width is determined
by the discontinuity of $\Pi(q^2)$ across the positive real
$q^2$-axis. To calculate $F(m^2/s)$ according to Eq.~(\ref{NLchange}),
we need to consider the contribution to $\Pi(q^2)$ from the graphs in
Fig.~\ref{ourgraphs}, where the quark in the outer loop is massless
and the quark in the inner loop has mass $m$. Call this contribution
$\Pi_G(q^2,m)$. Then
\begin{equation}
{ 1 \over 2\pi i}
\left[\Pi_G(s+i\epsilon,m) - \Pi_G(s-i\epsilon,m)\right]  =
{ 1 \over 12\pi^2}\
\left({ \alpha_s \over \pi}\right)^2
\left[
-0.115295 + F(m^2/s)
\right].
\label{discPi}
\end{equation}
Here the normalization factor $1/(12 \pi)$ is determined by the Born
graph. As dictated by Eq.~(\ref{NLchange}), we understand the graphs
to include their renormalization subtractions in the $\overline{\rm
MS}$ scheme. In particular, we subtract the pole part of the heavy
quark loop.

On the right hand side of Eq.~(\ref{discPi}), the value of the graphs
for a massless quark is represented by the constant term. The mass
dependence is contained in $F(m^2/s)$, where we have defined $F(0) =
0$. It is useful to  extract $F(m^2/s)$, by subtracting the $m=0$
graphs:
\begin{eqnarray}
\lefteqn{
{ 1 \over 12\pi^2}\
\left({ \alpha_s \over \pi}\right)^2
 F(m^2/s) =}
\label{FfromPi}\\
&&{ 1 \over 2\pi i}
\left[\Pi_G(s+i\epsilon,m) - \Pi_G(s-i\epsilon,m)\right]
-{ 1 \over 2\pi i}
\left[\Pi_G(s+i\epsilon,0) - \Pi_G(s-i\epsilon,0)\right].
\nonumber
\end{eqnarray}
The advantage is that if we take the difference inside the integrals
then the loop integrations are convergent. We write the three-loop
diagrams of Fig.~\ref{ourgraphs}, subtracted at zero mass, in the
form of two loop effective diagrams as shown in Fig.~\ref{effdiag}.
The double wavy line is the effective gluon propagator. It is
constructed from the gluon propagator with a quark loop insertion,
\begin{equation}
{ i \over k^2 + i\epsilon}\,
\left(
-g^{\mu\nu} + { k^\mu k^\nu \over k^2}
\right)
{\cal P}(k^2,m^2)\,,
\label{gluonprop}
\end{equation}
by replacing ${\cal P}(k^2,m^2)$ by ${\cal P}(k^2,m^2) - {\cal
P}(k^2,0)$. We use
\begin{equation}
{\cal P}(k^2,m^2)-{\cal P}(k^2,0) =
{ 2 m^2 \alpha_s \over \pi}
\int_{0}^{1} d\alpha \int_{0}^{1}d\beta\
{1 \over k^2 - M(\alpha,\beta)^2 + i \epsilon}\,,
\end{equation}
where $M(\alpha,\beta)^2=m^2\beta/[\alpha (1-\alpha)]$. We write the
diagrams of  Fig.~\ref{effdiag} as integrals over Feynman parameters
and integrate analytically over as many of the Feynman parameters as
possible. This involves a certain amount of computer algebra, for
which we use FORM \cite{FORM}. The structure of the result is simple
enough that we can take the discontinuity indicated in
Eq.~(\ref{FfromPi}) analytically. Then we perform the remaining
Feynman parameter integrals numerically by Monte Carlo integration.
(We use VEGAS \cite{vegas}.)

\begin{figure}[htb]
\centerline{ \DESepsf(effective.epsf width 10 cm) }
\caption{The two loop effective diagrams.}
\label{effdiag}
\end{figure}

Given $F(m^2/s)$, one can calculate $G(m^2/s)$ from Eq.~(\ref{FtoG}).
However, this procedure loses numerical accuracy for large $m^2/s$,
where the small number $G(m^2/s)$ is obtained from Eq.~(\ref{FtoG})
as the difference of two large numbers. A better procedure it to
calculate $G(m^2/s)$ directly. An added advantage of calculating
$G(m^2/s)$ directly is that Eq.~(\ref{FtoG}) then provides a check
on both calculations.

To calculate $G(m^2/s)$ directly, we use the Collins-Wilczek-Zee
(CWZ) renormalization prescription \cite{CWZ}. For the sake of
definiteness, let us say that the heavy quark with mass $m$ is the top
quark. The CWZ prescription is to renormalize the ultraviolet
divergences for all subgraphs that do {\it not} contain top quark
lines according to the $\overline{\rm MS}$ prescription. Subgraphs that
{\it do} contain top quark lines are renormalized by subtraction at
zero momentum. Then the value of a renormalized graph with external
particles consisting of light quarks and gluons and with external
momenta that are small compared to $m$ is the $\overline{\rm MS}$
value if the graph has no top quark loops, and zero in the large $m$
limit if the graph does have top quark loops. Thus $\alpha_s$ in this
scheme is to be identified with  the five-flavor $\overline{\rm MS}$
version of $\alpha_s$. Now in Eq.~(\ref{R2}) or
Eq.~(\ref{NLchange}), the function $G(m^2/s)$ is defined to be the
function that reflects the effect of top quark loops on $R_Z$ at the
three loop order of perturbation theory, but this means perturbation
theory in the five-flavor $\overline{\rm MS}$ version of $\alpha_s$.

Thus to calculate $G(m^2/s)$ we have only to calculate the graphs of
Fig.~\ref{ourgraphs} in the CWZ renormalization scheme. For these
graphs, all of the divergent subgraphs contain the heavy quark loop,
so all of the subtractions are at zero momentum. Since no
$\overline{\rm MS}$ subtractions are needed, it suffices to perform
the entire calculation in exactly four dimensions. From these graphs,
we construct $G(m^2/s)$ using
\begin{equation}
{ 1 \over 12\pi^2}\
\left({ \alpha_s \over \pi}\right)^2
 G(m^2/s) =
{ 1 \over 2\pi i}
\left[\Pi_G^{\rm CWZ}(s+i\epsilon,m)
- \Pi_G^{\rm CWZ}(s-i\epsilon,m)\right].
\label{GfromPi}
\end{equation}
which is the analogue of Eq.~(\ref{FfromPi}).

As with the calculation of $F(m^2/s)$, we can calculate the three
loop graphs as two loop graphs with an effective gluon propagator,
as shown in Fig.~\ref{effdiag}. In this case the effective gluon
propagator is given by Eq.~(\ref{gluonprop}) with ${\cal
P}(k^2,m^2)$ replaced by ${\cal P}(k^2,m^2) - {\cal P}(0,m^2)$,
reflecting the CWZ zero-momentum subtraction.  We represent this as
\begin{equation}
{\cal P}(k^2,m^2)-{\cal P}(0,m^2) =
- {\alpha_s \over \pi}
\int_{0}^{1} d\alpha\
{ \alpha^2\, (1 - \alpha^2/3) \over 1-\alpha^2}
{k^2 \over k^2 - M(\alpha)^2 + i \epsilon}\,,
\label{PCWZ}
\end{equation}
where $M(\alpha)^2 = 4 m^2/(1-\alpha^2)$. Some zero-momentum
subtractions are called for in the effective graphs of
Fig.~\ref{effdiag}. We need not worry about the overall subtraction,
since it is a constant, independent of $q^2$, and hence its
discontinuity across the positive $q^2$ axis vanishes. There are,
however, one loop subgraphs in Fig.~\ref{effdiag} that are divergent
by power counting and should be subtracted at zero momentum.
In the second and third graphs, there are divergent quark
self-energy subgraphs of the form $A(k^2)\,k\cdot \gamma$. According to
the CWZ prescription, these are replaced by $[A(k^2)-A(0)]\,k\cdot
\gamma$. In the first graph of Fig.~\ref{effdiag}, there is a
surprise: the counter-terms vanish when calculated
in exactly four dimensions because they have a factor of zero arising
from the numerator algebra. Thus the integral corresponding to this
diagram is finite.

As with the calculation of $F(m^2/s)$, we write the diagrams of
Fig.~\ref{effdiag} as integrals over Feynman parameters. We
take the discontinuity indicated in Eq.~(\ref{GfromPi}) numerically,
by choosing a small but finite value for $\epsilon$. Then we perform
the Feynman parameter integrals numerically by Monte
Carlo integration. (We checked that the result does not depend
on $\epsilon$ to within our integration errors.)

In Table~\ref{results} we present our numerical results for $F(m^2/s)$
and $G(m^2/s)$ for various values of quark mass $m$ at $\sqrt{s} =
M_{Z} = 91.188 GeV$. The integration error is $\lesssim 5$ in the last
digit. Using these data, we verify (to within the integration errors)
Eq.~(\ref{FtoG}), which relates $F$ and $G$.

\begin{table}
\caption{Calculated values of $F(m^2/s)$ and $G(m^2/s)$.}
\smallskip
\begin{tabular}{|l|l|l|||l|l|l|}
$m$(GeV)    & $F(m^2/M_Z^{2})$ & $G(m^2/M_Z^{2})$ &
 $m$(GeV)    & $F(m^2/M_Z^{2})$ & $G(m^2/M_Z^{2})$ \\ \hline
\ \ $m_c=1.3$\ \ \ \ \
&  0.00000031      &      &
  100        &  0.20160          & 0.0553     \\
\ \ $m_b=4.72$  &  0.0000392        & 0.869         &
  110        &  0.22572          & 0.0478     \\
\ \   10        &  0.000563         & 0.621         &
  120        &  0.24867          & 0.0418     \\
\ \   15        &  0.002199         & 0.487         &
  130        &  0.2703          & 0.0367      \\
\ \   20        &  0.00556          & 0.395         &
  140        &  0.2909          & 0.0325      \\
\ \   30        &  0.01854          & 0.2731        &
  150        &  0.31056          & 0.0293     \\
\ \   40        &  0.03933          & 0.1983        &
  170        &   0.3467          &  0.0238     \\
\ \   50        &  0.06523          & 0.1496        &
  180        &  0.3637          & 0.0216      \\
\ \   60        &  0.09335          & 0.1173        &
  200        &  0.3957          & 0.01825     \\
\ \   70        &  0.12174          & 0.0943        &
  250        &  0.4639          & 0.01256     \\
\ \   80        &  0.14953          & 0.0772        &
  600        &  0.7456         & 0.00277      \\
\ \ $M_Z$       &  0.1793           & 0.0635        &
 1000       &           & 0.001118            \\
\end{tabular}
\label{results}
\end{table}

\begin{figure}[htb]
\centerline{ \DESepsf(FandG.epsf width 10 cm) }
\caption{Calculated values of $F(m^2/s)$ and $G(m^2/s)$
versus $m$ for $\protect\sqrt s = 91.188$ GeV.}
\label{resultsplot}
\end{figure}

These results are plotted in Fig.~\ref{resultsplot}. For small
$m^2/s$, $F(m^2/s)$ is the physically relevant function in
Eq.~(\ref{R2}). We see from the table or the figure that $F$ vanishes
in this limit. A reasonable (2\%) approximation for $0.05 \lesssim
m/\sqrt s \lesssim 0.3$ is
\begin{equation}
F(m^2/s) \approx { m^4 \over s^2} \times \left\{
-0.474894
+ \log(s/m^2)
+{ m \over \sqrt s}  \left[
-0.5324
-0.0185 \log(s/m^2) \right]
\right\}.
\end{equation}
The first two terms here are taken from the recent evaluation of
Chetyrkin \cite{chetnew}. We confirm these coefficients within the
limits of our numerical calculation.  The next terms result from a
numerical fit to our data.

For large $m^2/s$, $G(m^2/s)$ is the physically relevant function. We
see that $G$ vanishes in this limit.  That is, the heavy quark
decouples \cite{DecTheor}. A reasonable (1\%) approximation
for $m/\sqrt s \gtrsim 1$ is
\begin{equation}
G(m^2/s) \approx { s \over m^2}\times\left\{
  {44 \over 675}
+ {2 \over 135} \log(m^2/s)
-{\sqrt{s} \over m}
\left[0.001226
+0.001129\log(m^2/s)
\right]
\right\}.
\label{Glargem}
\end{equation}
The first two terms here are taken from a recent evaluation of
Chetyrkin \cite{Chet}. We confirm these coefficients within the limits
of our numerical calculation.  The remaining terms result from a
numerical fit to our data. The fact that these terms are small
indicates that the large mass expansion works well. The present
calculation thus serves as a direct test confirming the usefulness of
the methods of Euclidean asymptotic expansions of Feynman integrals
(see \cite{TkaCh} and  references therein).

In summary, this calculation completes in a certain sense the
evaluation of the top quark mass dependence of $O(\alpha_s^2)$
corrections to the hadronic decay width of $Z$-boson. We have found
that quark mass dependence can be evaluated rather simply, without
depending on expansions in $m^2/s$ or $s/m^2$, in diagrams of a fairly
high order. We expect that the results may be useful for the analysis
of $\Gamma(\tau^{-} \rightarrow {\rm hadrons})$.

\acknowledgements
We are grateful to J.\ Collins, R.\ Haydock,  T.\ Hebbeker, B.\ Lampe,
and D.\ Strom for helpful comments and conversations.  This work was
supported by the U.S. Department of Energy under grant No.
DE-FG06-85ER-40224.

\end{document}